\documentclass{iopart}
\usepackage[utf8]{inputenc}
\usepackage{xcolor}
\usepackage{graphicx}
\usepackage[normalem]{ulem}
\usepackage[noend]{algorithm2e}
\usepackage{url}

\newcommand{\cfit}{$C_{\rm fit}$~}
\newcommand{\cshift}{$C_{\rm shift}$~}
\newcommand{\cangle}{$C_{\rm angle}$~}
\newcommand{\cage}{$C_{\rm age}$~}

\newcommand{\matchrate}{$3.5\%~$}
\newcommand{\removed}[1]{}
\newcommand{\added}[1]{#1}
\begin{document}

\title{A scalable system to measure contrail formation on a per-flight basis}

\author{Scott Geraedts$^{1}$, Erica Brand$^1$, Thomas R. Dean$^2$, Sebastian Eastham$^3$, Carl Elkin$^1$, Zebediah Engberg$^2$, Ulrike Hager$^1$, Ian Langmore$^1$, Kevin McCloskey$^1$, Joe Yue-Hei Ng$^1$, John C. Platt$^1$, Tharun Sankar$^1$, Aaron Sarna$^1$, Marc Shapiro$^2$, Nita Goyal$^1$}

\address{$^1$ Google, Mountain View, California}
\address{$^2$Breakthrough Energy, Kirkland, Washington}
\address{$^3$Laboratory for Aviation and the Environment, Massachusetts Institute of Technology}
\ead{geraedts@google.com}

\begin{abstract}
Persistent contrails make up a large fraction of aviation’s contribution to global warming. We describe a scalable, automated detection and matching (ADM) system to determine from satellite data whether a flight has made a persistent contrail. The ADM system compares flight segments to contrails detected by a computer vision algorithm running on images from the GOES-16 Advanced Baseline Imager. We develop a flight matching algorithm and use it to label each flight segment as a match or non-match. We perform this analysis on 1.6 million flight segments. The result is an analysis of which flights make persistent contrails several orders of magnitude larger than any previous work. We assess the agreement between our labels and available prediction models based on weather forecasts.
    Shifting air traffic to avoid regions of contrail formation has been proposed as a possible mitigation with the potential for very low cost/ton-CO2e.
    Our findings suggest that imperfections in these prediction models increase this cost/ton by about an order of magnitude. Contrail avoidance is a cost-effective climate change mitigation even with this factor taken into account, but our results quantify the need for more accurate contrail prediction methods and establish a benchmark for future development.
\end{abstract}


\section{Introduction}
\label{introduction}

Persistent contrails are cirrus clouds formed by aircraft as they fly through the upper atmosphere. Like all cirrus clouds, this `aviation-induced cirrus' both blocks outgoing long-wave infrared radiation and reflects incoming short-wave radiation \cite{myhre2001, burkhardt2011}. Over the past several years, the atmospheric science community has realized that the net effect of persistent contrails on the radiative balance of the planet is warming, by some measures more than the warming due to the carbon dioxide emissions of the aviation industry 
\cite{burkhardt2011, bock2016, ChenGettelman2013, Schumann2015dehydration, bickel2020estimating, lee2021contribution}. 
Aircraft only form persistent (i.e.~lasting longer than a few minutes) contrails when flying through pockets of air that are cold enough to satisfy the Schmidt-Appleman criteria \cite{schmidt1941, Appleman1953TheFO, schumann1996conditions} and have relative humidity greater than $100\%$ with respect to ice, so-called ice supersaturated regions (ISSR). ISSR are relatively rare and small, so the flight trajectory changes needed to avoid contrail formation are also small \cite{avila2019reducing, teoh2020mitigating, teoh2022aviation}. Therefore, adopting a contrail avoidance approach of avoiding flying through ISSR could significantly reduce the warming impact of the aviation industry at potentially small cost. This is one of the most cost-effective climate change mitigations available \cite{caldiera2021}. 

Evaluating the effectiveness of contrail avoidance is difficult without empirical observation. Observing enough contrails to do large-scale evaluation was previously a difficult problem, but the development of contrail detection machine learning models based on satellite imagery \cite{meijer2022contrail, Ng2023} has made it possible to automatically observe very large numbers of contrails with much higher accuracy than earlier approaches.

In this work we use historical infrared images from the GOES-16 geostationary satellite to detect persistent contrails.
Based on the distance between these contrails and recorded flight paths, we classify all flights as either
making or not making contrails. Our method is fully automated and can be scaled to assess all flights over a wide area, for example this work covers an area including the entire contiguous United States over an aggregate period of 168 hours. We analyze properties of the observed contrails such as their age, and dependence on flight density and time of day/year, similar to previous works \cite{duda2004, vazqueznavarro2015} but with orders of magnitude more data.

Contrail avoidance requires the ability to predict which flights will make contrails. There has been considerable progress on developing models that can predict contrail formation \cite{schumann2012cocip, schumann2020cocip2, teoh2022aviation, fritz2020,yin2022}. However there has been no large-scale attempt to assess how accurately these models predict contrail formation on a per-flight basis. Multiple works \cite{gierens2020, Agarwal_2022} have shown that the critically important input of humidity in the upper atmosphere is often poorly predicted by weather forecasts. 

Existing efforts to assess the cost and benefits of contrail avoidance \cite{avila2019reducing, teoh2020mitigating, teoh2022aviation, caldiera2021} assume perfect contrail prediction, and though this is not a prerequisite for adopting contrail avoidance, imperfections will both decrease the benefits (since some contrails will not be predicted in advance and therefore not avoided) and increase the cost (since some flights will spend extra fuel attempting to avoid creating contrails when their original flight path would not have passed through any ISSR). 

In this work we compare our observations to the output of contrail prediction models. For each flight segment, we ask the models whether or not a contrail would form, check the GOES-16 ABI imagery to see if an observed contrail matches the flight segment, and tally up precision and recall for each model. Though our contrail observations are not perfect, we expect that the GOES-16 ABI instrument, with its $2 km$ resolution and sensors viewing the parts of the electromagnetic spectrum where contrails are expected to trap the most heat, should be able to detect most contrails with a strong warming impact.

Our work is the first attempt to assess, for a large number of flights and independent of modeled humidity data, whether each flight made a contrail. As such it allows us to compare the performance of different contrail prediction models and the weather data that they use. It also allows us to estimate how imperfections on prediction models affect the cost of contrail avoidance.

The remainder if this work is organized as follows. In Sec.~\ref{sec:data} we describe the data we use. In Sec.~\ref{flight_matching} we specify the flight matching algorithm that determines whether a given flight segment matches an observed contrail. In Sec.~\ref{prediction_models} we outline the different prediction models that we will compare our observations to. In Sec.~\ref{sec:matching_results} we describe the properties of the contrail-flight matches we observe, such as their age and their distribution in space and time. In Sec.~\ref{sec:prediction_results} we compare our observed matches to the results of contrail prediction models.

\section{Materials and Methods}
\label{sec:methods}
\subsection{Data}
\label{sec:data}
We start by automatically detecting contrails using computer vision methods described in Ng et al.\cite{Ng2023}, which uses infrared images from the GOES-16 Advanced Baseline Imager (ABI)\cite{goodman2019goes} as input. These images cover much of the Western hemisphere. The images have a temporal resolution of 10 minutes and a spatial resolution of approximately 2 km.

We derive flight trajectories from ground-based ADS-B data provided by FlightAware (\url{https://flightaware.com}). All flight paths are resampled so that there is one flight waypoint per minute. Flights are divided into 10-minute `flight segments', the length of 10 minutes was chosen to yield segment sizes of $\approx 150km$, the typical flight length inside ISSR\cite{gierens2000size}. This leads to 10 waypoints total for most segments, though some segments have fewer waypoints, e.g.~at the start and end of the flight and where there are gaps in ADS-B coverage. We include in our analysis all segments with at least 6 waypoints. The ADSB data contain a small number \added{($0.3\%$)} of \removed{unrealistic}\added{incorrect} flight trajectories whose\removed{. To remove these, the} velocity between \removed{each pair of} waypoints \removed{are computed and the flight segment is dropped if this velocity} exceeds 1,234.8 $km/h$ (Mach 1 at sea level).\added{We drop these segments because no civilian aircraft present in the ADS-B data can fly at that speed}. Additionally, only flight waypoints whose altitude is above 7000 meters are included since persistent contrail conditions are very rare below this altitude. Apart from these considerations we process all flight waypoints available from FlightAware.

Weather data in this work comes from the European Center for Medium-range Weather Forecasts (ECWMF). We use both high-resolution (HRES) forecast and ERA5 reanalysis data\cite{hersbach2020era5}. The forecast data is obtained on a $0.1^\circ\times0.1^\circ$ grid at model altitude levels while the reanalysis data is obtained on a $0.25^\circ\times0.25^\circ$ grid. In order to study the effects of vertical resolution reanalysis data is obtained on both model levels (which have $\approx 10 hPa$ resolution) and pressure levels (which have 25-50 hPa resolution). When weather forecasts are used we use the 
forecast initialized at midnight UTC on the relevant day, which would be 0-23 hours old at the time of the flight waypoint.

\begin{figure}
    \centering
    \includegraphics[width=0.8\textwidth]{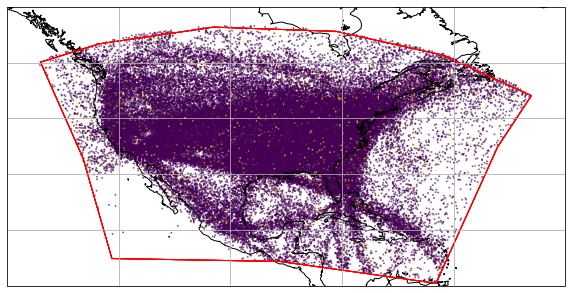}
    \caption{Illustration of the region (red) considered in this work. Each point in the figure represents one of the advected flight segments considered in this work. Yellow points correspond to segments that matched contrails while purple points correspond to segments that do not match contrails. The points in the figure represent $1\%$ of the data considered in this work.}
    \label{regionfigure}
\end{figure}

In this work we analyze all flights inside the region shown in Fig.~\ref{regionfigure} containing the entire contiguous USA as well as much of the rest of North America. 
\removed{The data analysed in this paper comes from the time-period Apr 4 2019 -- Apr 4 2020.}  We analyze data sampled from across a whole year to account for seasonality of contrail formation\cite{meijer2022contrail}, \added{starting on Apr 4 2019, just after GOES-16 ABI began taking images every 10 minutes. Our data therefore runs to Apr 4 2020.}
In Mar 2020 there was a large drop in air traffic density due to the COVID-19 pandemic, but we have found that quantities such as the fraction of flights matching contrails and the skill of prediction models were not meaningfully different in that month. 
We focus on 168 hours worth of data, distributed in 28 6-hour chunks, uniformly sampled by month and at different times of day, since this is more computationally tractable than processing an entire year.

\subsection{Flight Matching algorithm}
\label{flight_matching}

Flight matching compares, for each flight, the position of observed contrails to the position a contrail would be at  if the flight had made a contrail.
The expected position at the time that the GOES-16 ABI imaged a scene is determined through three-dimensional advection of the flight waypoints using the third-order Runge-Kutta method \cite{bogacki19893} and winds taken from weather data.
Each waypoint is advected for two hours, covering the next 11 GOES-16 ABI images (10 minute time-steps). 
Because ice crystals in a contrail fall over time, we also sink the waypoints vertically. The crystals fall at terminal velocity, assuming a crystal size obtained by performing a quadratic fit to the distribution of ice crystal sizes given by CoCiP\cite{schumann2012cocip}. The resulting function for radius $r$ is:
\begin{equation}
r = e^{1.48 + 0.61 \log(t)  + 0.08 \log(t)^2}.
\end{equation}
Where $r$ is in $\mu m$ and $t$ is in hours. \added{Since the amount of fall in two hours is small compared to the weather data’s resolution, assuming a fixed crystal size gives very similar results. The advantage of modeling the crystal growth is that it allows the same method to be used with longer advections in future works.} We also descend the waypoints by $50 m$ at the start of advection to account for wake vortex downwash. The value of $50 m$ is similar to the values used for the initial sinking in CoCiP.\cite{schumann2012cocip}
We next assess whether the expected location of each flight is close to any observed contrails, using a method adapted from Duda et al.\cite{duda2004}. Given a contrail detected by our computer vision algorithm, we rotate the contrail and the advected flight path to a rotated coordinate system indexed by the coordinates $v$ and $w$. In this coordinate system the contrail runs from $v=-L/2$ to $v=L/2$, with $L$ being the length of the contrail, and has $w=0$. The advected flight waypoints have coordinates $(w_i, v_i)$. We consider advected flight waypoints that overlap with the contrail (between $v=-L/2 - \Delta$ and $v=L/2 + \Delta$, with $\Delta=\sqrt{C_{\rm shift} + C_{\rm fit}}$). The flight and the contrail match only if there are at least two overlapping waypoints.

We then find another coordinate transformation
\begin{eqnarray}
w \rightarrow (w + W)\cos(\theta) + (v + V)\sin(\theta) \nonumber\\
v \rightarrow (v+V)\cos(\theta) - (w + W)\sin(\theta), \nonumber
\end{eqnarray}
i.e. we shift the coordinates by $(W, V)$ and also rotate by the angle $\theta$. The coordinate transformation minimizes the cost function:
\begin{eqnarray}
    {\rm match~error} &=& [C_{\rm fit} \frac{1}{N}\sum_{i=1}^{N} w_i^2 + C_{\rm shift} (V^2+W^2) \nonumber\\
    &&+ C_{\rm angle} (1 - cos(\theta)) + C_{\rm age}].
\label{eq:match_score}
\end{eqnarray}
A low match error indicates a good match. In other words, we shift and rotate the coordinates to find a coordinate system where the advected flight waypoints line up exactly with the contrail. The `fit' term quantifies how successful we were at doing this, while the `shift' and `angle' quantify how big a shift and rotation is required in order to get the waypoints and the contrail to line up. A big shift and/or rotation implies that the advected flight and contrail weren't that close to begin with, and this leads to a high match error. 

The \cfit term represents residual linearity error - advected flight waypoints that are not themselves linear are unlikely to have created a linear contrail. Our flight matching therefore penalizes curved flight trajectories, and in fact $0.03\%$ of flight segments are so non-linear that this term will prevent them from matching any contrail.

The \cshift term is dominated by the uncertainty in the wind. If this wind is incorrect, the correct flight may advect to the wrong location, and these errors get larger the longer we advect for. We compare the wind forecast data used for advection with the wind data produced by the Mode-S data broadcast by airplanes\cite{dehann2011}. The Mode-S system uses flight transponders and ground-based radar to obtain the wind velocity for selected aircraft with a high degree of accuracy. Mode-S data was obtained from FlightAware for 326,000 waypoints over the contiguous United States on Aug 20, 2021. We found a root mean squared error of $11$ km/h. 

The \cangle term is dominated by the difference in wind errors at different locations, which will rotate the advected flight path. For the same waypoints as used in the \cshift analysis, we compared the wind speeds at two locations separated by the length of our typical segment ($150$ km), finding a root mean squared error of $3.8$ degrees/hour. 

The constants \cfit, \cshift and \cangle were chosen so that each term should be $\approx1$ when the error is as large as the mean error. Specifically:
\begin{eqnarray}
    && C_{\rm fit} = \frac{1}{(1 km)^2} \\
    && C_{\rm shift} = \frac{1}{(10 km/h \times t)^2} \\
    && C_{\rm angle} = \frac{1}{1 - \cos(5~degrees/hour \times t)}
\end{eqnarray}
where $t$ is the difference between the time of the flight and the time the contrail was observed, in hours. 
Since a typical error will lead to each term in Eq.~(\ref{eq:match_score}) being $1$, a flight is labeled a match if the overall match error is $\le 3$.

The \cage term is a correction factor for the \cshift and \cangle terms which become increasingly permissive as a contrail gets older. It was tuned by a small grid search after the other terms had been selected, to yield an age distribution of contrails similar to what has been found in previous observation studies, in particular Figure 6 of Vazquez-Navarro et al.\cite{vazqueznavarro2015}:
\begin{equation}
    C_{\rm age} = t.
    \label{cage}
\end{equation}

Additionally, we have attempted to handle the case where multiple flights match what appears to be the same contrail. 
Such a case could happen if two contrails are so close together that they are detected as only one contrail, (in which case calling them both a match is correct) or if two flights have similar locations but different altitudes (in which case only one should be called a match). We have attempted to balance these concerns by calling
both flights a match, unless one of the matching flights is much better than the other (specifically, the difference in match errors $\ge1$). In that case the worse matching flight is not counted as matching that contrail (it can still match other contrails).

Matching individual segments to contrails leads to problems when contrails overlap the segment boundaries. Therefore during flight matching we match contrails to entire flights, and if a match is found label all segments that overlap the contrail as matches. A pseudocode version of the algorithm is given in Algorithm \ref{pseudocode}.
\makeatletter
\newcommand{\SetKwMetaData}[4]{%
  \algocf@newcommand{@#1}[1]{\DataSty{#2#3}\ArgSty{##1}\DataSty{#4}}%
  \algocf@newcommand{#1}{%
    \@ifnextchar\bgroup{\csname @#1\endcsname}{\DataSty{#2}\xspace}}%
  }%
\makeatother
  
\begin{algorithm}
\caption{Pseudocode for flight matching algorithm}
\label{pseudocode}

\SetKwData{Flight}{Flight}
\SetKwData{Contrail}{Contrail}
\SetKwData{AllFlights}{AllFlights}
\SetKwData{AllContrails}{AllContrails}
\SetKwData{Segments}{Segments}
\SetKwData{BetterMatchFound}{BetterMatchFound}
\SetKwMetaData{Scores}{Scores}{[}{]}
\SetKwData{Segment}{Segment}
\SetKwMetaData{SegmentMatches}{SegmentMatches}{[}{]}
\SetKwData{OtherFlight}{OtherFlight}
\SetKwFunction{GetScore}{GetScore}
\SetKwFunction{GetSegments}{GetSegments}
\SetKwFunction{Overlaps}{Overlaps}

\SegmentMatches{all segments} $\leftarrow$ False

\For{\Flight in \AllFlights}{
    \For{\Contrail in \AllContrails}{
        \Scores{\Flight,\Contrail} $\leftarrow$ \GetScore{\Flight, \Contrail}
    }
}
\For{\Flight in \AllFlights}{
    \For{\Contrail in \AllContrails}{
        \If {\Scores{\Flight,\Contrail} $>$ 3}{continue}
        \BetterMatchFound $\leftarrow$ False
        
        \For{\OtherFlight in \AllFlights}{
            \If{\Scores{\Flight,\Contrail} $-$ \Scores{\OtherFlight,\Contrail} $>1$}{
                \BetterMatchFound $\leftarrow$ True
            }
        }
        \If{not \BetterMatchFound}{
            \For{\Segment in \GetSegments{\Flight}}{
                \If{\Overlaps{\Segment, \Contrail}}{
                    \SegmentMatches{\Segment} $\leftarrow$ True
                }
            }
        }    
    }
}

\end{algorithm}

\begin{figure}
    \centering
    \includegraphics[width=0.8\textwidth]{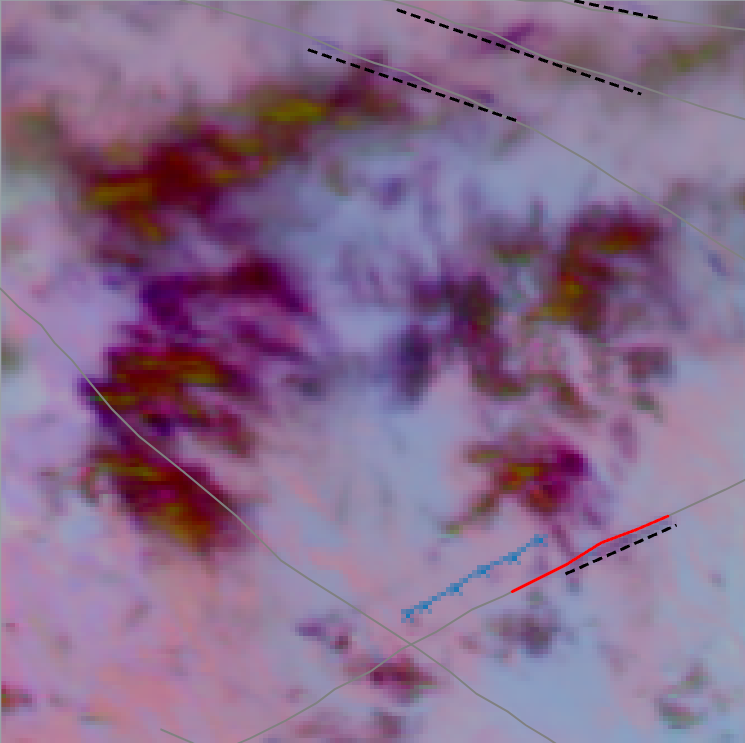}
    \caption{An example of our ADM system. The \added{four} dashed black lines indicate linearized contrails from our computer vision model. The red line indicates the advected flight path at the time the satellite image was taken and the blue line is the original flight path. We compare the red line with all black lines, and determine whether we think the flight segment in question matches a contrail (in this case, the result is that it does). The gray lines are all the advected flight paths that pass through the image.}
    \label{flight_matching_intro_fig}
\end{figure}

To assess the performance of flight matching we created a tool that displays sequences of images similar to Fig.~\ref{flight_matching_intro_fig}. Each image contains the flight segment in question, all other nearby flights, a false-color GOES-16 ABI infrared image, and detected contrails. We randomly selected 1000 flight segments from our dataset and three authors of this work assessed whether the flight segment matched a contrail, with a majority vote among the humans determining the correct label. We found that $88\%$ of the flights that the humans thought matched contrails were also labeled as matches by our automated matching. Only $50\%$ of the flights labeled as matches by the automated matching were labeled as matches by the humans. Most of the errors come from cases where there are a large number of flights, and multiple flights that are near the same contrail. The humans were able to uniquely determine which flight matched the contrail, while our automated matching was more likely to match all nearby flights to the same contrail. In the results section we study how the errors in the automated matching affect its results by showing results for both the automated matching and the data with human-evaluated flight matches.

\subsection{Prediction models}
\label{prediction_models}
We predict whether each flight segment will make a contrail using the methods described below. In the Results section we assess whether these predictions agree with our observations.

\subsubsection{Baseline}

The minimum requirements for persistent contrail formation are that the Schmidt-Appleman criterion (SAC)\cite{schmidt1941, Appleman1953TheFO, schumann1996conditions} is satisfied and the relative humidity over ice (RHi) is greater than $100\%$\cite{gierens2012ice}. The `Baseline' model evaluated in this work makes predictions solely based on these requirements. 
In order to account for subgrid scale variations and biases of RHi in weather forecast data, when predicting contrails it is common to apply a threshold for RHi different from $100\%$, or to rescale the data.\cite{schumann2012cocip, Schumann2021, teoh2022aviation, li2023}
To account for this we experiment with varying the RHi threshold when displaying results. When making predictions for an entire flight segment consisting of multiple waypoints, 
we multiply RHi by the binary SAC values and average the results before comparing them to the threshold (other ways of aggregating give similar or worse results).
 When computing SAC we assume an aircraft engine efficiency of 0.3 for all aircraft\cite{Ponater2002}, though trying other values (0.2, 0.4) does not have a noticeable effect.

For predicting contrail formation, weather variables are linearly interpolated from gridded historical weather data. RHi is computed from specific humidity which was logarithmically interpolated in altitude to handle it's large variations. We show results for both forecast and reanalysis weather data. The Baseline model only considers the weather at the flight waypoint timestamp.

\subsubsection{CoCiP}
The Contrail Cirrus Prediction model (CoCiP) \cite{schumann2012cocip, schumann2020cocip2} is a widely-used parameterized physics model of contrail evolution. After determining that a contrail will form and persist, CoCiP models contrail lifetime through initial downdraft, advection and fall, continually reassessing whether the contrail persists or sublimates. Modeling the microphysical properties of the contrail allows CoCiP to predict quantities such as optical depth and radiative forcing. 

Contrails which are too thin or short-lived to be visible in the GOES -16 ABI imagery, or which are obscured by clouds, may be predicted by the Baseline model to form contrails, and this will lead to disagreements between the Baseline model and our observations if the observations miss such hard-to-detect contrails. Since CoCiP tracks these quantities directly, it may be better able to predict which flights form contrails visible to GOES-16 ABI.  \added{See section \ref{sec:prediction_results} for details on how CoCiP was used to predict contrails visible to GOES-16 ABI.}
\removed{
Though CoCiP does not predict GOES-16 ABI visibility directly there are a few quantities we could use as a proxy.
For example, CoCiP predicts how long each contrail will persist, and thresholding based on this quantity might work since contrails with longer lifetimes will have a chance to appear in more GOES-16 ABI images. One drawback of using contrail persistence time as a proxy for observability is that it does not account for small or faint contrails. A better proxy might therefore be the product of optical depth and width, integrated over contrail lifetime. Even this doesn't capture the case where CoCiP predicts that a contrail will be difficult to observe because of other clouds above or below it. The proxy for observability we use is the integral of the predicted radiative forcing of the contrail in the long wave infrared, for the times for which we are making observations. Much like RHi in the Baseline model, we average this `long-wave energy forcing' (LWEF) across segments and show results for different thresholds. LWEF is the quantity most similar to how different the contrail pixels in GOES-16 ABI images are from the surrounding pixels. It is smaller for contrails which are small, optically thin, short-lived, or obscured by clouds, and so by focusing on cases where CoCiP predicts a high LWEF we are focusing on the contrails which should have high contrast with surrounding pixels and therefore be among the easiest to detect in GOES-16 ABI longwave imagery.}

We obtain CoCiP predictions through the API made available by Breakthrough Energy at \url{https://api.contrails.org}, which implements the original CoCiP algorithm along with modifications developed in \cite{Schumann2015dehydration, teoh2020mitigating, teoh2022aviation, breakthrough_docs}.

\subsection{Metrics}
\label{metrics}
We define `precision' as the fraction of predicted flight segments which are considered to match a contrail using our method, and `recall' as the fraction of matched contrails which are successfully predicted. A prediction model which always agrees with our observations would have precision=1 and recall=1.

The precision and recall of a given model depends on what threshold of relative humidity (or LWEF for CoCiP) is used. 
 A high threshold can avoid predicting observed contrails that aren't there, but at the cost of potentially missing some observed contrails (high precision, low recall), while a low threshold will correctly find most observed contrails but also potentially predict that some flights will make contrails when no contrail is observed (high recall, low precision). Which threshold is appropriate depends on the application. We compute results for multiple different thresholds to show all the different results that different thresholds can produce.

Previous works\cite{teoh2020mitigating, caldiera2021} have compared the climate benefits of avoiding contrails with the cost of the extra fuel needed to do the avoidance. They assume a perfect contrail prediction model, but imperfections in prediction models will change the cost/benefit analysis. Roughly speaking, the benefit needs to be multiplied by the recall, because some contrails are never predicted by the model and never avoided.  The cost also needs to be increased by 1/precision, because some flights are rerouted (using extra fuel) for no benefit. \removed{In total the cost/benefit penalty factor (CBPF) is 1/(precision $\times$ recall).} \added{In total the cost/benefit is increased by 1/(precision x recall), which we call the cost-benefit penalty factor (CBPF).} This CBPF is an estimate only since its exact value depends on the details of contrail avoidance.

\section{Results and Discussion}

\subsection{Properties of observed flight matches}
\label{sec:matching_results}
In this work we analyze using our automated detection and matching (ADM) system 255,341 flights, broken into $\sim1.8$ million flight segments. We find that \matchrate  of flight segments in FlightAware are observed to match a contrail, and $14.5\%$ of flights have at least one segment that matches a contrail. 

When contrails initially form they are $\approx 100 m$ in width\cite{schumann2012cocip} which makes them nearly impossible to see in infrared GOES-16 ABI images which have $2~km$ resolution at nadir. Wind shear and diffusion spread the contrail out as it ages.\cite{schumann2012cocip}
The age of each contrail when it is first observed is shown in Fig.~\ref{match_age}. Most flight segments start matching contrails about half an hour after formation, with the mean time until first observation being 41 minutes, approximately consistent with existing literature \cite{duda2004, vazqueznavarro2015, Chevallier2023}.
There are likely many flight segments that make contrails which sublimate before they are large enough to be seen in GOES-16 ABI. We label those flight segments non-matches since we never detect those contrails. This is acceptable since such contrails have only a small climate impact, and contrails which form in ISSR are expected to have lifetimes much longer than our minimum detection time \cite{schumann2012cocip}.

Fig.~\ref{match_age} also shows the effect of incorporating time (as in Eq.~\ref{cage}) in the flight matching algorithm, which prevents some very old flight segments from matching contrails. With or without the $C_{\rm age}$ term, the number of contrails with initial detection age of two hours is small, which motivates our decision to only compare flights to contrails detected in the first two hours after the flight. It is possible that a contrail could spread out so slowly that it is only large enough to be detected (and therefore matched) after two hours, in which case we would miss that match in this work, but based on Fig.~\ref{match_age} we think the number of such contrails should be very small. This is consistent with previous works \cite{freudenthaler1995, jensen1998, duda2004, schumann2012cocip}. We have tried extending this system past two hours of advection and studying the resulting matches, and based on multi-temporal visualizations of advected flight paths aligned with GOES-16 ABI imagery, matches of purported contrails first seeming to appear after 2 hours appear to be erroneous (data not shown).
\begin{figure}
    \centering
    \includegraphics[width=0.8\textwidth]{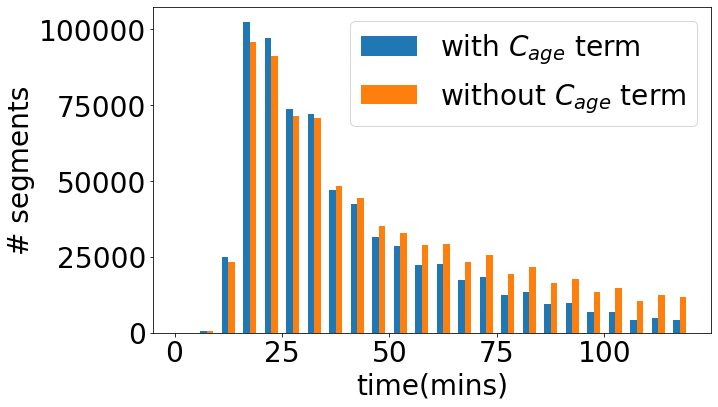}
    \caption{Histogram of the age of each flight segment at the first time it matches a contrail.}
    \label{match_age}
\end{figure}

\begin{figure}
    \centering
    \includegraphics[width=0.8\textwidth]{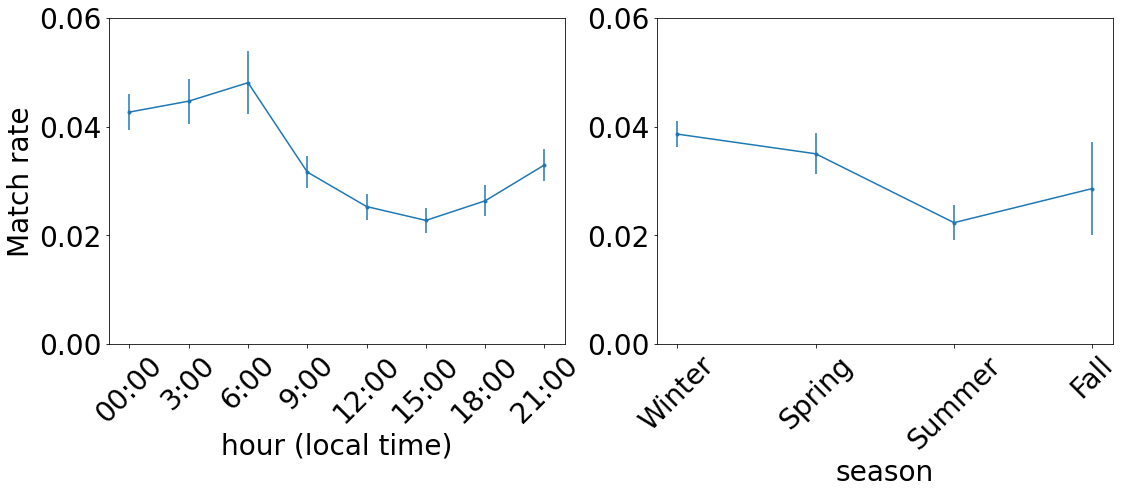}
    \caption{Match rate (fraction of flight segments observed to match a contrail) as a function of time of day(left), and season (right). Error bars are the standard error from averaging over different days.}
    \label{hour}
\end{figure}

Figure \ref{hour} shows the match rate (the fraction of flight segments the ADM system matches to a contrail) as a function of local time of day and season. The results seem to show a diurnal cycle with a higher fraction of flight segments being matched to contrails in the night and morning, and a lower fraction in the afternoon. This is consistent with previous works \cite{minnis1997, palikonda2005, minnis2013b, meijer2022contrail} which also found fewer contrails in the afternoon.  
Previous work~\cite{duda2004, palikonda2005, minnis2013a, meijer2022contrail} has suggested that fewer contrails form in the summer months, \added{which has important practical implications for contrail mitigation, and} appears to be supported by our data. 
To avoid creating correlations between time of day and time of year, the results in Fig.~\ref{hour} use 28 24-hour chunks (rather than 6 hour chunks).

\begin{figure}
    \centering
    \includegraphics[width=0.8\textwidth]{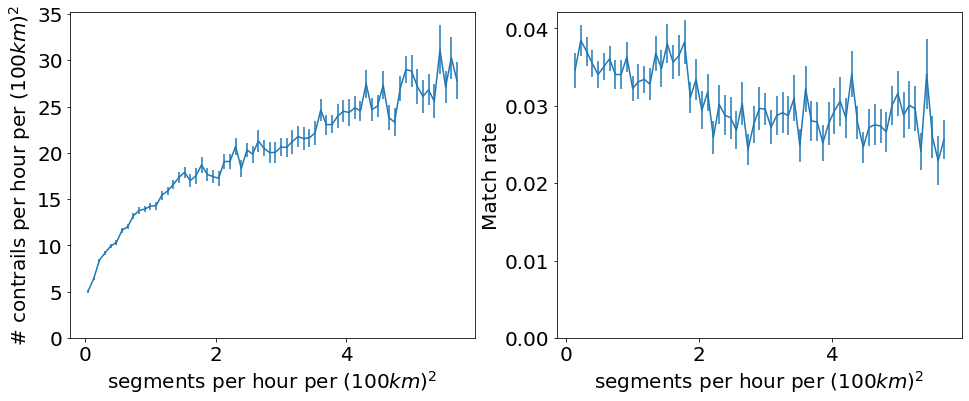}
    \caption{Number of contrails (left) and match rate (right) as a function of flight segment density. Fewer contrails are detected and matched at high flight densities. Error bars are one standard deviation, obtained by bootstrapping (resampling with replacement)}
    \label{flight_density}
\end{figure}

In Fig.~\ref{flight_density} we study the dependence of our results on flight density.
We divide our region into ($1^\circ$ latitude) $\times~(1^\circ$ longitude) $\times~(1$ hour) boxes, and compute the number of flights, contrails, and matches in each box. We then add together all the bins with similar numbers of flight segments, and compute the overall number of contrails detected and match rate. The result is a breakdown of our dataset by flight density. 

We find sublinear growth in the number of contrails with increased flight density (and a correspondingly lower match rate) consistent with previous findings \cite{bickel2020estimating,minnis2013b, duda2023}.
One explanation for this is that contrails are harder to detect when there are many of them overlapping, as suggested by Minnis et al.\cite{minnis2013b}. Another possibility is that in areas of high flight density there may not be enough excess water vapor available for all flights to make contrails \cite{Schumann2015dehydration}.

\subsection{Comparison with prediction models}
\label{sec:prediction_results}
We now compare our detection results to various prediction models. To illustrate our method, let us first consider the example of the Baseline model with HRES forecast weather data, and assuming that all flights with RHi$>95\%$ are predicted to make a contrail. The confusion matrix for this case is given in Table~\ref{confusion_matrix}. The results show that only a third of the contrails we observe are predicted, and no contrail is observed for $85\%$ of the segments which are predicted to make a contrail.

\begin{table}
\centering
\begin{tabular}{|c|c|c|}
\hline
  & Detected   & Not detected \\
\hline
Predicted     &  20,877 & 116,435 \\
Not predicted & 43,024 & 1,662,780 \\
\hline
Precision & \multicolumn{2}{c|}{0.15}\\
Recall & \multicolumn{2}{c|}{0.33} \\
\hline
\end{tabular}
\caption{Confusion matrix for the forecast Baseline model with an RHi $>95\%$ cutoff.}
\label{confusion_matrix}
\end{table}

As discussed above, the threshold of humidity $95\%$ is not the only possibility. We repeat the above study for different thresholds from $70-115\%$, and plot all the precision/recall results as points on a curve in Fig.~\ref{pr1}. We then do the same thing for the Baseline model with ERA5 reanalysis weather data defined on model vertical levels (ERA5$_{\rm model}$), and again for the Baseline Model with ERA5 data on pressure levels (ERA5$_{\rm pressure}$), and the CoCiP model using HRES forecast data. 

\added{Though CoCiP does not predict GOES-16 ABI visibility directly there are a few quantities we could use as a proxy.
For example, CoCiP predicts how long each contrail will persist, and contrails with longer lifetimes will have a chance to appear in more GOES-16 ABI images. One drawback of using contrail persistence time as a proxy for observability is that it does not account for small or faint contrails. A better proxy might therefore be the product of optical depth and width, integrated over contrail lifetime. Even this doesn't capture the case where CoCiP predicts that a contrail will be difficult to observe because of other clouds above or below it. The proxy for observability we use is the integral of the predicted radiative forcing of the contrail in the long wave infrared, for the times for which we are making observations. Much like RHi in the Baseline model, we average this `long-wave energy forcing' (LWEF) across segments and show results for different thresholds. LWEF is the quantity most similar to how different the contrail pixels in GOES-16 ABI images are from the surrounding pixels. It is smaller for contrails which are small, optically thin, short-lived, or obscured by clouds, and so by focusing on cases where CoCiP predicts a high LWEF we are focusing on the contrails which should have high contrast with surrounding pixels and therefore be among the easiest to detect in GOES-16 ABI longwave imagery.}

\removed{In the CoCiP case we threshold on LWEF rather than RHi, and} \added{We} use thresholds \added{over the range} $20,000 - 500,000 J/m$ \added{on the CoCiP-predicted LWEF}. We have also computed the performance of the CoCiP model using ERA5 reanalysis inputs, the results (not shown) were very similar to the Baseline model using reanalysis inputs.

\begin{figure}
    \centering
    \includegraphics[width=\textwidth]{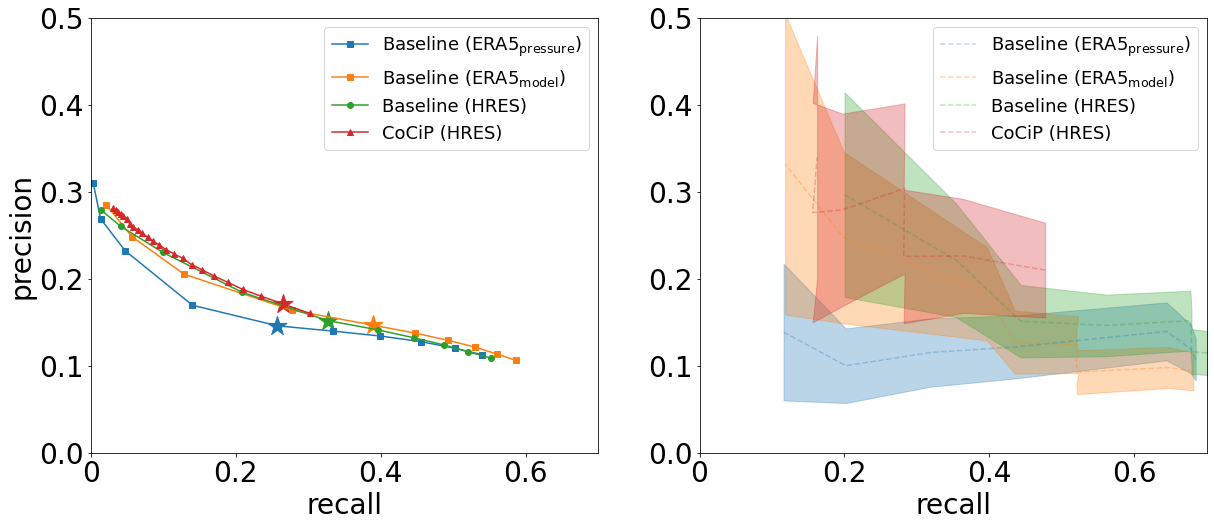}
    \caption{(Left) Precision/recall for the prediction models studied in this work, using our ADM system. We trace out curves by trying multiple different thresholds as described in the text. For Baseline models the starred points correspond to a threshold of $95\%$ RHi, and each other point results from adjusting the threshold by $5$ percentage points. CoCiP is similar but with $40,000 J/m$ for the starred point and steps of $20,000 J/m$. (Right) Precision/recall when comparing predictions to human flight matching results instead of automated matching. The prediction models agree better with the human labels, though the results are much noisier due to small sample sizes. Error bars represent one standard deviation computed using bootstrapping with 1000 samples. }
    \label{pr1}
\end{figure}

Comparing the different prediction models, the Baseline model based ERA5$_{\rm pressure}$ performs the worst. For example with a $95\%$ RHi threshold the model achieves precision 0.14, recall 0.25. This is not surprising: ISSRs are vertically thin and so it is not unexpected that a model with poor vertical resolution should have a hard time predicting them. The remaining models all give very similar performance. In particular the reanalysis and forecast data have very similar results when they both use the same vertical resolution.

The performance of CoCiP and the Baseline model were very similar when both used the same weather as input. In the plot we show CoCiP using LWEF as a proxy for whether we can detect a contrail. Using CoCiP predictions of lifetime or optical depth instead gives worse agreement with our observations.
Since thresholding on the CoCiP LWEF filters out the contrails predicted to be small, short-lived or cloud-obscured, its lack of improved precision over the Baseline model is notable. It suggests that the disagreement between our observations and the baseline model is not the difficulty in observing such contrails, but that the primary source of error in contrail forecasts comes from inaccurate RHi available from weather data.
Note that microphysical modeling is still useful to determine properties of contrails (such as radiative forcing) when RHi data is accurate. but we find no evidence that it improves predictions of whether contrails will be observed.

Note that neither the Baseline or CoCiP model handles interactions between contrails, and neither model predicts the drop in contrail formation at high flight density observed in Fig.~\ref{flight_density}. This may also contribute to the disagreement between model predictions and observations.

In these results we use the forecast wind data for advection, using reanalysis wind data the quantitative results are very similar and the relative ordering of the different weather models are unchanged.

The results above compare prediction models to observations, but our observations are not perfect. We now attempt to quantify the impact of these imperfections on the results in Fig.~\ref{pr1}.

The performance of the automated contrail detector in Ng et al.\cite{Ng2023} was evaluated relative to human-generated labels. It was found to miss $30\%$ of contrails and $30\%$ of the objects it recognized as contrails were in fact false positives. 
Errors of the contrail detector may lead to errors in the ADM system, or they may not, e.g.~if the detector incorrectly detects a contrail that is far from any flight paths, then no flight segment will be affected. This makes it difficult to exactly quantify the impact of contrail detection errors on our results, but due to such errors, even a perfect prediction model would achieve a maximum precision/recall in Fig.~\ref{pr1} around $0.7$.

To quantify the effects of automated flight matching errors on our results, in Fig.~\ref{pr1} we also compute precision and recall for our prediction models using only the flight segments with human-evaluated matching obtained above. Since for the human labeled data we have only 1000 flight segments (of which 29 were labeled by humans as matching contrails), the statistical error bars are much larger. To quantify this we bootstrap by resampling the data with replacement, the resulting error bars are indicated by the shaded regions in Fig.~\ref{pr1}. The results allow us to quantify how much an improved flight matching algorithm could improve the agreement between prediction models and our observations. We see that though the agreement is improved substantially the observed precision and recall falls well short of what a perfect prediction model could achieve. This is consistent with existing literature, for example Gierens et al.\cite{gierens2020} compared ERA5 reanalysis to aircraft based MOZAIC measurements and found $16\%$ precision and $21\%$ recall when assessing whether RHi$>100\%$, while Agarwal et al.\cite{Agarwal_2022} compared ERA5 measurements to radiosonde data and found that ERA5 incorrectly predicted the conditions for persistent contrail formation $87\%$ of the time.

\subsection{Implications for contrail avoidance}

In the example above of the forecast Baseline model with a $95\%$ RHi threshold, if contrail avoidance were attempted the number of flight segments we observe making contrails would be reduced by $33\%$, and of the flights whose flight paths were changed to avoid contrails, $85\%$ would not have matched a contrail even if they were not rerouted.
 This implies a CBPF of $1/(0.15 \times 0.33) \approx 20$. We can decrease the CBPF by choosing a better threshold, and it is possible to find a threshold where it is $\approx 16$ for all models. ($105\%, 115\%, 110\%$ RHi for ERA5$_{\rm pressure}$, ERA5$_{\rm model}$ and HRES, respectively, $20,000J/m$ for CoCiP) Furthermore we know that imperfections in our ADM system are artificially inflating the CBPF. If instead we used the mean values of the bootstrapped human labels as a guide, the CBPF is $8-13$.

\subsection{Discussion}

In this work we have matched a large number of flights over a wide area to observed contrails. Our method can be used to generate large datasets of contrails for further analysis, and could become the basis for a system of verifying contrail avoidance.

Our results establish a benchmark for the performance of contrail prediction models while we hope aiding in improving those models. An important step would be to improve forecasts of relative humidity at high altitudes. Alternatively, it may be possible to use our data to analyze and correct for possible biases in weather forecast data, or to use real-time observations of contrails to predict future contrail formation. Our method provides a way to measure whether approaches like these actually improve prediction accuracy. 

The values of the CBPF quoted above may also be artificially high due to errors in our detection model. Other factors (such as the cost of mistakenly diverting aircraft {\it into} contrail formation regions) are also not considered here. A more detailed estimation of the CBPF would be a useful direction for future work. Caldeira et al.\cite{caldiera2021} reports that contrail avoidance has benefits 1000 times larger than its cost, so even with the CBPF reported here contrail avoidance using current technology is likely a high value climate change mitigation strategy. 

Further improvements to the ADM system described here are possible.
Improvements to contrail detection are discussed in Ng et al.~\cite{Ng2023}.
The flight matching procedure can also be improved, especially in high flight density areas. 
A more detailed treatment of the case where multiple flights match the same contrail is desirable but difficult. An improved ground-truth flight-matching dataset would enable future progress. Such a dataset would not need to be as large as the one in this work (perhaps hundreds of flights, rather than hundreds of thousands), but would need to observe contrails closer to their formation time and confirm persistence by tracking them for a few hours. Ground-based cameras might be one way to build such a dataset.

Additional data inputs to flight matching algorithms could improve their accuracy: e.g., contrail altitude estimates could be compared to flight altitudes. Cloud top height can be extracted from geostationary images  \cite{strandgren2017cips}, but such models suffer from poor performance for thin clouds \cite{strandgren2017eval} so the models should be validated against contrails specifically to determine whether they are appropriate for this use case. Dealing with non-linear contrails and flight trajectories is another area for further improvement.
This work's flight matching algorithm also treats each detected contrail independently, but contrails appearing near each other in consecutive frames are likely the same contrail, and could be required to match the same flight segment, as done recently in Chevallier et al.\cite{Chevallier2023}. A dataset of tracked contrails, such as created on a small scale in Vazquez-Navarro et al.\cite{vazqueznavarro2010}, could lead to further improvement.

Our work found similar performance between reanalysis and same-day weather forecasts. Same-day weather forecasts are likely sufficient for dispatcher-, pilot- or air traffic control-led contrail avoidance, but if longer lead times are required for flight planning purposes performance may decrease. The methods used in this work could be used to quantify the size of that decrease.

This work provides an empirical method to assess whether a flight made a persistent contrail, but not all persistent contrails produce the same amount of warming. A useful extension of this work would be to observe the radiative forcing of each contrail and compare that to predictions.

This work establishes an empirical basis for evaluation of contrail avoidance strategies, beginning with the continental United States. The techniques which we demonstrate using the GOES-16 ABI can be readily extended to cover any area of the world with sufficiently high-resolution geostationary satellite coverage.

The dataset used to generate the results in this work is available at \url{https://storage.googleapis.com/contrails_measurement_paper_data/dataset.parquet.gzip}.
\newline
\ack
The authors thank Christopher Van Arsdale, Vincent Meijer, Maria Barbosa and Louis Robion for useful discussions, Marc Stettler and Anna Michalek for feedback on the manuscript, Ethan Koenig for contributions to our codebase and Alex Merose, Aaron Bell and the Google Anthromet team for helping to provide weather data. Dr. Eastham's participation in this research was partially supported by the U.S. Federal Aviation Administration Office of Environment and Energy through ASCENT, the FAA Center of Excellence for Alternative Jet Fuels and the Environment, Project 78 through FAA Award Number 13-C-AJFE-MIT under the supervision of Nicole Didyk-Wells. Any opinions, findings, conclusions or recommendations expressed in this material are those of the authors and do not necessarily reflect the views of the FAA.

\bibliography{ref}

\providecommand{\newblock}{}
\begin{thebibliography}{10}
\expandafter\ifx\csname url\endcsname\relax
  \def\url#1{{\tt #1}}\fi
\expandafter\ifx\csname urlprefix\endcsname\relax\def\urlprefix{URL }\fi
\providecommand{\eprint}[2][]{\url{#2}}

\bibitem{myhre2001}
Myhre G and Stordal F 2001 {\em Geophys. Res. Lett.\/} {\bf 28} 3119--3122

\bibitem{burkhardt2011}
Burkhardt U and Karcher B 2011 {\em Nature Climate Change\/} {\bf 1} 54--58

\bibitem{bock2016}
Bock L and Burkhardt U 2016 {\em Journal of Geophysical Research:
  Atmospheres\/} {\bf 121} 9717--9736 (\textit{Preprint}
  \eprint{https://agupubs.onlinelibrary.wiley.com/doi/pdf/10.1002/2016JD025112})
  \urlprefix\url{https://agupubs.onlinelibrary.wiley.com/doi/abs/10.1002/2016JD025112}

\bibitem{ChenGettelman2013}
Chen C~C and Gettelman A 2013 {\em Atmospheric Chemistry and Physics\/} {\bf
  13} 12525--12536
  \urlprefix\url{https://acp.copernicus.org/articles/13/12525/2013/}

\bibitem{Schumann2015dehydration}
Schumann U, Penner J~E, Chen Y, Zhou C and Graf K 2015 {\em Atmospheric
  Chemistry and Physics\/} {\bf 15} 11179--11199
  \urlprefix\url{https://acp.copernicus.org/articles/15/11179/2015/}

\bibitem{bickel2020estimating}
Bickel M, Ponater M, Bock L, Burkhardt U and Reineke S 2020 {\em Journal of
  Climate\/} {\bf 33} 1991--2005

\bibitem{lee2021contribution}
Lee D~S, Fahey D, Skowron A, Allen M, Burkhardt U, Chen Q, Doherty S, Freeman
  S, Forster P, Fuglestvedt J {\em et~al.\/} 2021 {\em Atmospheric
  Environment\/} {\bf 244} 117834

\bibitem{schmidt1941}
Schmidt E 1941 Die entstehung von eisnebel aus den auspuffgasen von flugmotoren
  eintrag von Ulrich Schumann, zur Sicherstellung des Zugangs zu diesem
  wissenschaftshistorisch wichtigen Dokument; mit Zustimmung des Rechteinhabers
  (Nachfolger des Oldenbourg Verlags)
  \urlprefix\url{https://elib.dlr.de/107948/}

\bibitem{Appleman1953TheFO}
Appleman H~S 1953 {\em Bulletin of the American Meteorological Society\/} {\bf
  34} 14--20

\bibitem{schumann1996conditions}
Schumann U 1996 {\em Meteorologische Zeitschrift\/} {\bf 5} 4--23

\bibitem{avila2019reducing}
Avila D, Sherry L and Thompson T 2019 {\em Transportation Research
  Interdisciplinary Perspectives\/} {\bf 2} 100033

\bibitem{teoh2020mitigating}
Teoh R, Schumann U, Majumdar A and Stettler M~E 2020 {\em Environmental Science
  \& Technology\/} {\bf 54} 2941--2950

\bibitem{teoh2022aviation}
Teoh R, Schumann U, Gryspeerdt E, Shapiro M, Molloy J, Koudis G, Voigt C and
  Stettler M~E~J 2022 {\em Atmospheric Chemistry and Physics\/} {\bf 22}
  10919--10935
  \urlprefix\url{https://acp.copernicus.org/articles/22/10919/2022/}

\bibitem{caldiera2021}
Caldeira K and McKay I 2021 {\em Nature\/} {\bf 593} 341--341
  \urlprefix\url{https://ideas.repec.org/a/nat/nature/v593y2021i7859d10.1038_d41586-021-01339-7.html}

\bibitem{meijer2022contrail}
Meijer V~R, Kulik L, Eastham S~D, Allroggen F, Speth R~L, Karaman S and Barrett
  S~R 2022 {\em Environmental Research Letters\/} {\bf 17} 034039

\bibitem{Ng2023}
Ng J~Y~H, McCloskey K, Cui J, Meijer V~R, Brand E, Sarna A, Goyal N, Arsdale
  C~V and Geraedts S 2023 Opencontrails: Benchmarking contrail detection on
  goes-16 abi (\textit{Preprint} \eprint{2304.02122})

\bibitem{duda2004}
Duda D~P, Minnis P, Nguyen L and Palikonda R 2004 {\em Journal of the
  Atmospheric Sciences\/} {\bf 61} 1132--1146

\bibitem{vazqueznavarro2015}
V\'azquez-Navarro M, Mannstein H and Kox S 2015 {\em Atmospheric Chemistry and
  Physics\/} {\bf 15} 8739--8749
  \urlprefix\url{https://acp.copernicus.org/articles/15/8739/2015/}

\bibitem{schumann2012cocip}
Schumann U 2012 {\em Geoscientific Model Development\/} {\bf 5} 543--580
  \urlprefix\url{https://gmd.copernicus.org/articles/5/543/2012/}

\bibitem{schumann2020cocip2}
Schumann U, Mayer B, Graf K and Mannstein H 2012 {\em Journal of Applied
  Meteorology and Climatology\/} {\bf 51} 1391 -- 1406
  \urlprefix\url{https://journals.ametsoc.org/view/journals/apme/51/7/jamc-d-11-0242.1.xml}

\bibitem{fritz2020}
Fritz T~M, Eastham S~D, Speth R~L and Barrett S~R~H 2020 {\em Atmospheric
  Chemistry and Physics\/} {\bf 20} 5697--5727
  \urlprefix\url{https://acp.copernicus.org/articles/20/5697/2020/}

\bibitem{yin2022}
Yin F, Grewe V, Castino F, Rao P, Matthes S, Dahlmann K, Dietm\"uller S,
  Fr\"omming C, Yamashita H, Peter P, Klingaman E, Shine K, L\"uhrs B and Linke
  F 2022 {\em Geoscientific Model Development Discussions\/} {\bf 2022} 1--34
  \urlprefix\url{https://gmd.copernicus.org/preprints/gmd-2022-220/}

\bibitem{gierens2020}
Gierens K, Matthes S and Rohs S 2020 {\em Aerospace\/} {\bf 7} ISSN 2226-4310
  \urlprefix\url{https://www.mdpi.com/2226-4310/7/12/169}

\bibitem{Agarwal_2022}
Agarwal A, Meijer V~R, Eastham S~D, Speth R~L and Barrett S~R~H 2022 {\em
  Environmental Research Letters\/} {\bf 17} 014045
  \urlprefix\url{https://dx.doi.org/10.1088/1748-9326/ac38d9}

\bibitem{goodman2019goes}
Goodman S~J, Schmit T~J, Daniels J and Redmon R~J 2019 {\em The GOES-R series:
  a new generation of geostationary environmental satellites\/} (Elsevier)

\bibitem{gierens2000size}
Gierens K and Spichtinger P 2000 On the size distribution of ice-supersaturated
  regions in the upper troposphere and lowermost stratosphere {\em Annales
  Geophysicae\/} vol~18 (Springer) pp 499--504

\bibitem{hersbach2020era5}
Hersbach H, Bell B, Berrisford P, Hirahara S, Hor{\'a}nyi A, Mu{\~n}oz-Sabater
  J, Nicolas J, Peubey C, Radu R, Schepers D {\em et~al.\/} 2020 {\em Quarterly
  Journal of the Royal Meteorological Society\/} {\bf 146} 1999--2049

\bibitem{bogacki19893}
Bogacki P and Shampine L~F 1989 {\em Applied Mathematics Letters\/} {\bf 2}
  321--325

\bibitem{dehann2011}
de~Haan S 2011 {\em Journal of Geophysical Research: Atmospheres\/} {\bf 116}
  (\textit{Preprint}
  \eprint{https://agupubs.onlinelibrary.wiley.com/doi/pdf/10.1029/2010JD015264})
  \urlprefix\url{https://agupubs.onlinelibrary.wiley.com/doi/abs/10.1029/2010JD015264}

\bibitem{gierens2012ice}
Gierens K, Spichtinger P and Schumann U 2012 {\em Atmospheric Physics:
  Background--Methods--Trends\/}  135--150

\bibitem{Schumann2021}
Schumann U, Poll I, Teoh R, Koelle R, Spinielli E, Molloy J, Koudis G~S,
  Baumann R, Bugliaro L, Stettler M and Voigt C 2021 {\em Atmospheric Chemistry
  and Physics\/} {\bf 21} 7429--7450
  \urlprefix\url{https://acp.copernicus.org/articles/21/7429/2021/}

\bibitem{li2023}
Li Y, Mahnke C, Rohs S, Bundke U, Spelten N, Dekoutsidis G, Gro{\ss} S, Voigt
  C, Schumann U, Petzold A and Kr\"amer M 2023 {\em Atmospheric Chemistry and
  Physics\/} {\bf 23} 2251--2271
  \urlprefix\url{https://acp.copernicus.org/articles/23/2251/2023/}

\bibitem{Ponater2002}
Ponater M, Marquart S and Sausen R 2002 {\em Journal of Geophysical Research:
  Atmospheres\/} {\bf 107} ACL 2--1--ACL 2--15 (\textit{Preprint}
  \eprint{https://agupubs.onlinelibrary.wiley.com/doi/pdf/10.1029/2001JD000429})
  \urlprefix\url{https://agupubs.onlinelibrary.wiley.com/doi/abs/10.1029/2001JD000429}

\bibitem{breakthrough_docs}
Shapiro M, Engberg Z, Teoh R, Stettler M, Dean T, Schemann U and Voigt C
  pycontrails: Python library for modeling aviation climate impacts
  \urlprefix\url{https://zenodo.org/record/8160906}

\bibitem{Chevallier2023}
Chevallier R, Shapiro M, Engberg Z, Soler M and Delahaye D 2023 {\em
  Aerospace\/} {\bf 10} ISSN 2226-4310
  \urlprefix\url{https://www.mdpi.com/2226-4310/10/7/578}

\bibitem{freudenthaler1995}
Freudenthaler V, Homburg F and Jäger H 1995 {\em Geophysical Research
  Letters\/} {\bf 22} 3501--3504 (\textit{Preprint}
  \eprint{https://agupubs.onlinelibrary.wiley.com/doi/pdf/10.1029/95GL03549})
  \urlprefix\url{https://agupubs.onlinelibrary.wiley.com/doi/abs/10.1029/95GL03549}

\bibitem{jensen1998}
Jensen E~J, Ackerman A~S, Stevens D~E, Toon O~B and Minnis P 1998 {\em Journal
  of Geophysical Research: Atmospheres\/} {\bf 103} 31557--31567
  (\textit{Preprint}
  \eprint{https://agupubs.onlinelibrary.wiley.com/doi/pdf/10.1029/98JD02594})
  \urlprefix\url{https://agupubs.onlinelibrary.wiley.com/doi/abs/10.1029/98JD02594}

\bibitem{minnis1997}
Minnis P, Kirk~Ayers J and Weaver S~P 1997 {\em 4 NASA Ref. Publ.\/} {\bf 1404}
  1157--1160

\bibitem{palikonda2005}
Palikonda R, Minnis P, Duda D~P and Mannstein H 2005 {\em Meteorologische
  Zeitschrift\/} {\bf 14} 525--536
  \urlprefix\url{http://dx.doi.org/10.1127/0941-2948/2005/0051}

\bibitem{minnis2013b}
Minnis P, Bedka S~T, Duda D~P, Bedka K~M, Chee T, Ayers J~K, Palikonda R,
  Spangenberg D~A, Khlopenkov K~V and Boeke R 2013 {\em Geophysical Research
  Letters\/} {\bf 40} 3220--3226 (\textit{Preprint}
  \eprint{https://agupubs.onlinelibrary.wiley.com/doi/pdf/10.1002/grl.50569})
  \urlprefix\url{https://agupubs.onlinelibrary.wiley.com/doi/abs/10.1002/grl.50569}

\bibitem{minnis2013a}
Duda D~P, Minnis P, Khlopenkov K, Chee T~L and Boeke R 2013 {\em Geophysical
  Research Letters\/} {\bf 40} 612--617 (\textit{Preprint}
  \eprint{https://agupubs.onlinelibrary.wiley.com/doi/pdf/10.1002/grl.50097})
  \urlprefix\url{https://agupubs.onlinelibrary.wiley.com/doi/abs/10.1002/grl.50097}

\bibitem{duda2023}
Duda D~P, Smith~Jr W~L, Bedka S, Spangenberg D, Chee T and Minnis P 2023 {\em
  Journal of Geophysical Research: Atmospheres\/} {\bf 128} e2022JD037554
  e2022JD037554 2022JD037554 (\textit{Preprint}
  \eprint{https://agupubs.onlinelibrary.wiley.com/doi/pdf/10.1029/2022JD037554})
  \urlprefix\url{https://agupubs.onlinelibrary.wiley.com/doi/abs/10.1029/2022JD037554}

\bibitem{strandgren2017cips}
Strandgren J, Bugliaro L, Sehnke F and Schr\"oder L 2017 {\em Atmospheric
  Measurement Techniques\/} {\bf 10} 3547--3573
  \urlprefix\url{https://amt.copernicus.org/articles/10/3547/2017/}

\bibitem{strandgren2017eval}
Strandgren J, Fricker J and Bugliaro L 2017 {\em Atmospheric Measurement
  Techniques\/} {\bf 10} 4317--4339
  \urlprefix\url{https://amt.copernicus.org/articles/10/4317/2017/}

\bibitem{vazqueznavarro2010}
Vazquez-Navarro M, Mannstein H and Mayer B 2010 {\em Atmospheric Measurement
  Techniques\/} {\bf 3} 1089--1101
  \urlprefix\url{https://amt.copernicus.org/articles/3/1089/2010/}

\end{thebibliography}

\clearpage
\end{document}